\documentclass{elsart}

\usepackage{amsmath,amssymb}

\begin{document}

\begin{frontmatter}
\bibliographystyle{elsart-num}

\title{Quantum Mechanics of Successive Measurements with Arbitrary Meter Coupling}

\author{Lars M. Johansen}
\address{Department of Technology, Buskerud University College,
N-3601 Kongsberg, Norway}

\ead{lars.m.johansen@hibu.no}

\author{Pier A. Mello}
\address{
Instituto de F\'{\i}sica, Universidad Nacional Aut\'{o}noma de M\'{e}xico,
M\'{e}xico, D.F. C.P. 04510}

\ead{mello@fisica.unam.mx}

\date{July 8, 2008}

\begin{abstract}

We study successive measurements of two observables using von Neumann's measurement model. The two-pointer correlation for arbitrary coupling strength allows retrieving the initial system state. We recover L\"uders rule,  the Wigner formula and the Kirkwood-Dirac distribution in the appropriate limits of the coupling strength.

\end{abstract}


\end{frontmatter}

\section{Introduction}

The time evolution of an isolated system obeys, in Quantum Mechanics (QM), the Schr\"odinger equation.
When an observable $\hat{A}$ is measured on the system, the ``standard rule" is that the only possible outcomes are the eigenvalues of $\hat{A}$, each result occurring with a probability given by Born's rule. The ``orthodox" view of QM asserts in addition that due to the measurement of
$
\hat{A} = {\sum}_n a_n \mathbb{P}_{a_n}
$
--with (possibly degenerate) eigenvalues $a_n$ and eigenprojectors $\mathbb{P}_{a_n}$--
a discontinuous change (von Neumann's postulate \cite{Neumann-MathFounQuanMech:55,Wigner-ProbMeas:63}, or L\"uder's rule \cite{Lueders-UEbeZust:51} if the spectrum is degenerate) governed by probability laws occurs in the state of the system $\rho$:
if $\hat{A}$ is measured disregarding the measurement outcome (a non-selective projective measurement \cite{Johansen-Quantheosuccproj:07}), $\rho$ changes to
\begin{equation}
\bar{\rho} = {\sum}_n \mathbb{P}_{a_n} \rho \; \mathbb{P}_{a_n}\; ;
\label{reduction_nonselective}
\end{equation}
if the outcome $a_{n}$ is selected (a selective projective measurement \cite{Johansen-Quantheosuccproj:07}),
$\rho$ changes to (the ``state projection", or ``collapse", postulate)
\begin{equation}
\rho ' = \frac{\mathbb{P}_{a_n} \rho \; \mathbb{P}_{a_n}}
{{\rm Tr}(\rho \; \mathbb{P}_{a_n})} \; .
\label{reduction_selective}
\end{equation}
Postulating the above discontinuous changes is really a way of avoiding inclusion of the instrument in the description of the measurement process. In particular, the collapse postulate, which is of no consequence as long as single measurements are performed, is relevant for successive measurements when the instrument is not included: if two not necessarily commuting observables, $\hat{A}$ and then $\hat{B}$, are measured in succession, the projection formula (\ref{reduction_selective}) gives the joint probability of finding $a_n$ and then $b_m$ as
\begin{equation}
W^W_{b_m  a_n} = {\rm Tr} (\rho \; \mathbb{P}_{a_n}\mathbb{P}_{b_m}
\mathbb{P}_{a_n}).
\label{wigner_rule}
\end{equation}
This is ``Wigner's formula" \cite{Wigner-ProbMeas:63}, first written down
for the case of nondegenerate observables.

In contrast, including the instrument, as in von Neumann's model (vNM)  \cite{Neumann-MathFounQuanMech:55,Bell+Nauenberg-MoraAspeQuanMech:66}, allows investigating the dynamical basis underlying the measurement process. In such an approach one obtains information on the system by observing some property of the instrument, like the pointer position $\hat{Q}$, for which QM can only make statistical predictions. Ref. \cite{Arthurs+Kelly-SimuMeasPairConj:65} generalized the vNM and considered two instruments for the simultaneous measurement of position and momentum. Ref.  \cite{Aharonov+AlbertETAL-ResuMeasCompSpin:88} analyzed successive measurements with the vNM and studied the average of the first observable for weak coupling with the meter (called the ``weak value"), conditioned on a post-selection. The vNM was also used in Refs. \cite{Peres-QuanLimiDeteWeak:89,Lamb-Sequmeasquanmech:87} to study a sequence of measurements, and in Ref. \cite{Hay+Peres-QuanClasDescMeas:98} to study the problem of von Neumann's cut.
More generally, the description of quantum measurements without inclusion of the Hilbert space of the instrument is represented by the theory of effects and operations \cite{Kraus-StatEffeOper:83}.
Ref. \cite{konrad} contains, among other topics, a discussion of successive measurements with arbitrary system-meter interaction strength, which the author relates to the theory of effects and operations; an important list of references on these points can also be found there.

In the present letter we employ the vNM to investigate further the problem of two successive measurements in QM. Our model is similar to that of Ref. \cite{Peres-QuanLimiDeteWeak:89}. We study the measurement process as a function of the strength of the coupling between the system and the pointers and, in particular, we analyze the correlation between the two pointers.

As a reminder to the reader, and in order to establish the notation, we consider in section \ref{sec:single} the von Neumann measurement of a single observable. Here, the focus is on the pointer position, which conveys information about the system. The pointer momentum is of no interest in this case. The pointer momentum commutes with the von Neumann interaction Hamiltonian, and hence remains unaffected by the interaction. Next, in section \ref{sec:successive}, we consider successive measurements. Our primary interest is in the correlation between the two pointers. As we will see, both the position and the momentum of the first pointer give useful information when the correlation with the second pointer is considered. This gives us two correlation functions. We show that these may be expressed as the real and imaginary part of a complex quasi-probability over the eigenvalues of the two observables.  This quasi-probability reduces to Wigner's formula (\ref{wigner_rule}) in the limit of a very strong coupling for the first measurement, without ever needing the state-projection formula (\ref{reduction_selective}). For very weak coupling the quasi-probability reduces to the  Kirkwood-Dirac distribution \cite{Kirkwood-QuanStatAlmoClas:33,Dirac-AnalBetwClasQuan:45}: our model can thus be regarded as giving a derivation of both results in the appropriate limits. We also obtain the reduced density matrix of the system after the first measurement for an arbitrary coupling strength, and show how one recovers L\"uder's rule (\ref{reduction_nonselective}) in the limit of strong coupling.

State reconstruction based on the successive measurement of two observables is a problem only rarely addressed in the literature (see, however, Ref.  \cite{Johansen-Quantheosuccproj:07}). In section \ref{sec:reconstruct} we show that the measurement scheme presented here permits the reconstruction of the initial system state -- using arbitrary coupling strengths.
This scheme requires a separate measurement of every pair of eigenprojectors belonging to the two observables. Furthermore, it requires measuring both the position \emph{and} the momentum of the first pointer. We find that the class of informationally complete observables is the same as was found in Ref. \cite{Johansen-Quantheosuccproj:07}, namely that the observables should be nondegenerate and complementary.

\section{Single measurements}
\label{sec:single}

 We first consider a von Neumann measurement of a single observable
$\hat{A} = \sum_n a_n \mathbb{P}_{a_n}$, where the eigenvalues $a_n$ are allowed to be degenerate. It will be instructive to contrast some of the results for single measurements with those for successive measurements.

We assume the system to be coupled to a pointer, whose position and momentum are represented by the Hermitian operators $\hat{Q}$ and $\hat{P}$.
The system-pointer interaction is taken to be
\cite{Neumann-MathFounQuanMech:55}
\begin{equation}
\hat{V}(t) = \epsilon \; \delta (t-t_1) \hat{A} \hat{P}\; , \hspace{5mm}  t_1 > 0\; ,
\label{V_single}
\end{equation}
with an {\em arbitrary} interaction strength  \cite{Peres-QuanLimiDeteWeak:89} $\epsilon$. We disregard the intrinsic evolution of the system and the pointer and assume that $V$ represents the full Hamiltonian.
The evolution operator is given by
\begin{equation}
\hat{U}(t)
= {\rm exp}[-i \int_0^t \hat{V}(t')dt']
= {\rm exp}[-i \epsilon \; \theta (t-t_1)\hat{A}\hat{P}] ,
\label{U_single}
\end{equation}
in units of $\hbar =1$; $\theta (\tau)$ is the step function. If the density matrix of the system plus the pointer at $t=0$ is
$
\rho = \rho_{s} \rho_{M}
$
($M$ stands for ``meter"),
for $t>t_1$ it is given by
\begin{equation}
\rho^{(\hat{A})}
= \sum_{n  n' }
\mathbb{P}_{a_n}
\rho_{s}
\mathbb{P}_{a_{n'}}
(e^{-i\epsilon a_n \hat{P}}
\rho_{M}
e^{i\epsilon a_{n'} \hat{P}})
\; .
\label{rho_t>t1 single}
\end{equation}
We now observe the pointer position $\hat{Q}$ to obtain
information on the system.
According to Born's rule, and for a pure state $| \chi \rangle$ for the pointer,
the $Q$ probability density for $t>t_1$ is given by
(notice that we are {\em not} using the projection postulate for the {\em state})
\begin{subequations}
\begin{eqnarray}
p^{(\hat{A})}(Q)
&=&\sum_n
W^{(\hat{A})}_{a_n} \;
|\chi(Q-\epsilon a_n)|^2 ,
\label{p(Q)}
\\
W^{(\hat{A})}_{a_n}&=&{\rm Tr}_s (\rho_{s} \mathbb{P}_{a_{n}}),
\label{Born_single_meas}
\end{eqnarray}
\label{p(Q),Born,single}
\end{subequations}
where $W^{(\hat{A})}_{a_n}$
is the Born probability for the result $a_n$ and
$|\chi(Q-\epsilon a_n)|^2$
is the original $Q$ probability density displaced by the amount
$\epsilon a_n$ (its width is $\sigma_Q$).
This result shows that the ``standard rule" of QM that the only possible outcomes of the measurement of an observable are the eigenvalues of the corresponding Hermitean operator, Born's rule giving their probability,
has to be translated into the
probability
of the pointer, which in turn ``mirrors" the former
only in the idealized limit of very strong coupling, $\epsilon/\sigma_Q \gg 1$
\cite{Neumann-MathFounQuanMech:55,Peres-QuanLimiDeteWeak:89}.
Notice that, in contrast,
we took for granted that it is for the probability of occurrence of $Q$ that we are entitled to apply Born's rule.
Should $\hat{Q}$ be microscopic
\footnote {For instance, in a Stern-Gerlach experiment designed to measure the $z$ projection of the electron spin, $(1/2)\sigma_z$, $\hat{Q}$ represents the $\hat{p}_z$ momentum component of the electron.},
we might need to measure it with a macroscopic instrument, observe its pointer position $\hat{Q}'$ and assume that we can apply Born's rule for the $Q'$ probability density.

However, it is remarkable that, for {\em arbitrary} $\epsilon$,
the average of $\hat{Q}$ in units of $\epsilon$ is given by
\begin{equation}
\langle \hat{Q} \rangle^{(\hat{A})}/ \epsilon
= \sum_n  a_n W^{(\hat{A})}(a_n)
= {\rm Tr} (\rho_{s}\hat{A}).
\label{<Q>}
\end{equation}
(The original $Q$ distribution is supposed centered at $Q=0$.) This is the Born average of the observable $\hat{A}$ in the original state of the system \cite{Aharonov+AlbertETAL-ResuMeasCompSpin:88}.

The particular case in which the observable $\hat{A}$ in
the interaction is replaced by the projector
$\mathbb{P}_{a_{\nu}}$ is of great importance.
We designate the eigenvalues of $\mathbb{P}_{a_{\nu}}$
by $\pi = 1,0$, and its eigenprojectors by
$\mathbb{P}_{a_{\nu}}^{\pi}$.
Then
$
\mathbb{P}_{a_{\nu}}^1 = \mathbb{P}_{a_{\nu}}
$,
$
\mathbb{P}_{a_{\nu}}^0 = I - \mathbb{P}_{a_{\nu}}
$.
For these eigenvalues and eigenprojectors, Eq. (\ref{<Q>}) gives
\begin{equation}
\langle \hat{Q}  \rangle^{(\mathbb{P}_{a_{\nu}})} / \epsilon
= {\rm Tr}_s (\rho_s \mathbb{P}_{a_{\nu}}^{\pi=1})
=W^{(\hat{A})}_{a_{\nu}} \; .
\label{<Q>proj 1}
\end{equation}
For successive measurements the situation will be more subtle.

From Eq. (\ref{rho_t>t1 single}) we now compute the reduced density operator of the system tracing over the pointer, to find
\begin{equation}
\rho_{s}^{(\hat{A})}
= \sum_{n  n' }
g_{n n'}^{(\hat{A})}(\epsilon)\;
\mathbb{P}_{a_n}
\rho_{s}
\mathbb{P}_{a_{n'}} \; .
\label{rho_s_t>t1 0}
\end{equation}
We have defined the characteristic function of the pointer momentum distribution
\begin{equation}
g^{(\hat{A})}(\epsilon(a_{n}-a_{n'}))
= {\rm Tr}[\rho_M e^{-i\epsilon(a_{n}-a_{n'})\hat{P}}]
\equiv g_{n n'}^{(\hat{A})}(\epsilon) ,
\label{g_nn'}
\end{equation}
which we shall call the {\em decoherence factor}
\cite{Peres-QuanLimiDeteWeak:89}.
As an example, if we assume the pure Gaussian state
\begin{equation}
\chi(Q)
=(2\pi\sigma^2_{Q})^{-1/4}
 \exp(-Q^2/4\sigma^2_{Q})
\label{chi,Gaussian}
\end{equation}
for the pointer, we find \cite{Peres-QuanLimiDeteWeak:89}
\begin{equation}
g_{n n'}^{(A)}(\epsilon)
= \exp[-(\epsilon^2 / 8\sigma_Q^2)(a_n - a_{n'})^2].
\label{g_nn'_Gauss}
\end{equation}
Result (\ref{rho_s_t>t1 0}) is valid for an arbitrary value of the coupling strength.
In the strong-coupling limit it reduces to L\"uders rule of Eq. (\ref{reduction_nonselective}),
originally postulated by L\"uders \cite{Lueders-UEbeZust:51},
derived later by assuming the measurement to be repeatable and minimally disturbing  \cite{Ludwig-Mess:53,Goldberger+Watson-MeasTimeCorrQuan:64}, and then given a dynamical derivation in Ref. \cite{Bell+Nauenberg-MoraAspeQuanMech:66} using vNM.

\section{Successive measurements}
\label{sec:successive}

We now turn to the problem of measuring two observables
in succession. We assume two pointers
(with momentum and coordinate operators $\hat{P}_i$, $\hat{Q}_i$)
whose interaction with the system
\begin{equation}
\hat{V} (t) = \epsilon_1 \delta (t-t_1) \hat{A} \hat{P}_1
+ \epsilon_2 \delta (t-t_2) \hat{B} \hat{P}_2, \;\;\;\;\;
0 < t_1 < t_2 ,
\label{V 2meas}
\end{equation}
is designed to measure the observable $\hat{A}$ defined above with the first pointer at time $t_1$, and the observable
$
\hat{B} = \sum_m b_m \mathbb{P}_{b_m}
$
with the second pointer at time $t_2$ (the $b_m$'s may also be degenerate).
The unitary evolution operator is given by
\begin{equation}
\hat{U}(t)
=\exp[-i\epsilon_2 \theta(t-t_2) \hat{B} \hat{P}_2]
\exp[-i\epsilon_1 \theta(t-t_1) \hat{A} \hat{P}_1].
\label{U 2meas}
\end{equation}
If the density operator describing the system plus the two pointers is, at $t=0$,
$
\rho =  \rho_{s} \rho_{M_1} \rho_{M_2}
$,
for $t>t_2$, i.e., after the second interaction, it is given by
\begin{eqnarray}
&&\rho^{(\hat{B} \leftarrow \hat{A})}= \sum_{nn'mm'}
(\mathbb{P}_{b_{m}} \mathbb{P}_{a_{n}}  \rho_{s} \; \mathbb{P}_{a_{n'}} \mathbb{P}_{b_{m'}})
\nonumber \\
&&
\hspace{3.5cm}
\cdot \left(
e^{-i\epsilon_1 a_n \hat{P_1}} \rho_{M_1}
e^{i\epsilon_1 a_{n'}\hat{P_1}} \right)
\left(
e^{-i\epsilon_2 b_m \hat{P_2}}  \rho_{M_2}
e^{i\epsilon_2 b_{m'} \hat{P_2}}
\right)\; .
\hspace{5mm}
\label{rho t1<t<t2 12 txt}
\end{eqnarray}

We now study what information we can obtain about the system
by observing the two pointer positions $\hat{Q}_1$ and $\hat{Q}_2$
for $t>t_2$.
From Eq. (\ref{rho t1<t<t2 12 txt}) we obtain, using Born's rule,
the $Q_1$, $Q_2$ joint probability density and,
when the pointers are prepared in pure Gaussian states $\chi^{(i)}(Q_i)$, $i=1,2$,
we find the correlation of the two pointer positions as
\begin{eqnarray}
\frac{\langle \hat{Q}_1 \hat{Q}_2 \rangle\
^{(\hat{B} \leftarrow \hat{A})}}{\epsilon_1 \epsilon_2}
&=& \Re  \sum_{nm} a_{n} b_{m}
W^{(\hat{B} \leftarrow \hat{A})}_{b_{m}a_{n}}(\epsilon_1),
\label{<Q1Q2> A 2}
\end{eqnarray}
where
\begin{eqnarray}
W^{(\hat{B} \leftarrow \hat{A})}_{b_{m}a_{n}}(\epsilon_1)
&=&\sum_{n'}
g_{nn'}^{(\hat{A})}(\epsilon_1)
Tr_s[\rho_{s}(\mathbb{P}_{a_{n'}} \mathbb{P}_{b_{m}} \mathbb{P}_{a_{n}})]
\hspace{10mm}
\label{W(b,a) B A}
\end{eqnarray}
is a {\em ``quasi-probability"} for $a_n$ and $b_m$. The reason for this naming convention will be explained shortly. 
The decoherence factor $g_{nn'}^{(\hat{A})}(\epsilon_1)$ is given in Eq. (\ref{g_nn'_Gauss}).
We note that the pointer correlation (in units of $\epsilon_1 \epsilon_2$) depends on the strength $\epsilon_1$ for the first measurement, but not on $\epsilon_2$.

We now consider again the same Hamiltonian of Eq. (\ref{V 2meas}) but,
after the second interaction has acted, i.e., for $t > t_2$, we observe,
on a second sub-ensemble,
the momentum $\hat{P}_1$ of the first pointer instead of its position, and the position $\hat{Q}_2$ of the second pointer. The resulting correlation between $\hat{P}_1$ and $\hat{Q}_2$ is
\begin{equation}
\frac{\langle \hat{P}_1 \hat{Q}_2 \rangle
 ^{(\hat{B} \leftarrow \hat{A})}}
{\epsilon_1 \epsilon_2}
 =\frac{1}{2\sigma^2_{Q_1}}
\Im \sum_{nm}  a_{n} b_{m}
W^{(\hat{B} \leftarrow \hat{A})}_{b_{m}a_{n}}(\epsilon_1)
\label{<P1Q2> A 2}
\end{equation}
where the quasi-probability
$W^{(B \leftarrow A)}_{b_{m}a_{n}}(\epsilon_1)$
is given in Eq. (\ref{W(b,a) B A}).

Eqs.  (\ref{<Q1Q2> A 2}) and (\ref{W(b,a) B A})  (which generalize to two successive measurements the result of Eqs. (\ref{<Q>}) and (\ref{Born_single_meas}) for a single measurement) are written in a manner similar to the correlation of two classical sets of variables $\{a_n\}$, $\{b_m\}$, with the exception that now $W^{(\hat{B} \leftarrow \hat{A})}_{b_{m}a_{n}}(\epsilon_1)$ is not necessarily real and non-negative, and we shall refer to it as the quasi-probability for the variables $a_n$, $b_m$, when the measurements of $\hat{A}$ and $\hat{B}$ are performed in succession. The quasiprobability $W^{(\hat{B} \leftarrow  \hat{A})}_{b_{m}a_{n}}(\epsilon_1)$ is a quantum generalization of the classical joint probability concept. It has the following properties:
\renewcommand{\theenumi}{\roman{enumi}}
\begin{enumerate}

\item

The marginal probabilities are
\begin{subequations}
\begin{eqnarray}
        \sum_m  W^{(\hat{B} \leftarrow
\hat{A})}_{b_{m}a_{n}}(\epsilon_1) &=& Tr(\rho_{s} \mathbb{P}_{a_n}),\\
        \sum_n  W^{(\hat{B} \leftarrow
\hat{A})}_{b_{m}a_{n}}(\epsilon_1) &=& Tr(\rho_{s}^{(\hat{A})} \mathbb{P}_{b_m}).
\end{eqnarray}
\end{subequations}
The first marginal coincides with the probability distribution of
$a_n$
on the system $\rho_s$ prior to the measurement. The second marginal coincides with the probability distribution of
$b_m$
after the measurement of $\hat{A}$, i.e. with respect to the reduced state $\rho_{s}^{(\hat{A})}$ given by Eq. (\ref{rho_s_t>t1 0}).

\item

In the strong-coupling limit, $\epsilon_1 \to \infty$, we have
\begin{equation}
W^{(\hat{B} \leftarrow \hat{A})}_{b_{m}a_{n}}(\epsilon_1)
\to
Tr_s[\rho_{s}(\mathbb{P}_{a_{n}} \mathbb{P} _{b_{m}}
\mathbb{P}_{a_{n}})]
\end{equation}
which is the joint probability given by Wigner's rule (\ref{wigner_rule}) (real and non-negative).
Thus the two-pointer correlation of Eq. (\ref{<Q1Q2> A 2}) reduces to
an average evaluated with a probability given by Wigner's formula.

\item

In the weak-coupling limit, $\epsilon_1 \to 0$, we have
\begin{equation}W^{(\hat{B} \leftarrow \hat{A})}_{b_{m}a_{n}}(\epsilon_1)
\to
K_{b_{m}a_{n}} \equiv Tr_s[\rho_{s}(\mathbb{P}_{b_{m}} \mathbb{P}_{a_{n}})],
\end{equation}
which is Kirkwood's quasi-probability (complex, in general).
The pointer correlation reduces to
\begin{subequations}
\begin{eqnarray}
\langle \hat{Q}_1 \hat{Q}_2 \rangle^{(\hat{B}\leftarrow \hat{A})}/ \epsilon_1
\epsilon_2
&=& Tr_s \Big[ \rho_{s}\frac12(\hat{B}\hat{A}+\hat{A}\hat{B}) \Big]
\hspace{15mm}
\label{corr weak coupl d}  \\
&=&\sum_{nm} a_n b_m W^{MH}_{b_{m} a_{n}} ,
\label{corr weak coupl e}
\\
W^{MH}_{b_{m} a_{n}}
&=&\frac12 Tr
\left[
\rho^{(s)}(\mathbb{P}_{b_{m}} \mathbb{P}_{a_{n}} + \mathbb{P}_{a_{n}} \mathbb{P}_{b_{m}})
\right] \; .
\label{MH}
\end{eqnarray}
\label{corr weak coupl MH}
\end{subequations}
$W^{MH}_{b_{m} a_{n}}$ is the real part of the Kirkwood quasi-probability distribution
\cite{Kirkwood-QuanStatAlmoClas:33,Dirac-AnalBetwClasQuan:45,Steinberg-Condprobquantheo:95,Johansen+Luis-NoncWeakMeas:04}, also known as Margenau-Hill's distribution \cite{Margenau+Hill-CorrbetwMeasQuan:61}.
The Margenau-Hill distribution may take negative values
and hence cannot be regarded as a joint probability in the classical sense \cite{Margenau+Hill-CorrbetwMeasQuan:61}.

\item

If the projectors $\mathbb{P}_{a_{n}}$, $\mathbb{P}_{b_{m}}$ appearing in Eq.  (\ref{W(b,a) B A}) commute, $[\mathbb{P}_{a_{n}}, \mathbb{P}_{b_{m}}] = 0, \;\;\forall n,m$, the quasiprobability reduces to
$W^{(\hat{B} \leftarrow \hat{A})}_{b_{m}a_{n}}(\epsilon_1)
=Tr_s[\rho_{s}(\mathbb{P}_{b_{m}} \mathbb{P}_{a_{n}})]
$
for arbitrary values of $\epsilon_1$;
this is the standard, real and non-negative, quantum-mechanical definition of the joint probability of $a_n$ and $b_m$ for commuting observables.
The correlation of the two pointer positions measured in units of $\epsilon_1 \epsilon_2$ coincides, for an arbitrary coupling strength  $\epsilon_1$, with the standard result for the correlation of the two observables $\hat{A}$ and $\hat{B}$, i.e. the r.h.s. of Eq. (\ref{<Q1Q2> A 2}) reduces to the form $Tr(\rho_{s}\hat{A}\hat{B})$.

\end{enumerate}

\section{State reconstruction}
\label{sec:reconstruct}

In this section we devise a state reconstruction scheme based on successive measurements. To this end, we shall consider the particular case in which the observables $\hat{A}$ and $\hat{B}$ appearing in the interaction (\ref{V 2meas}) are replaced by two new observables: the projectors $\mathbb{P}_{a_{\nu}}$ and $\mathbb{P}_{b_{\mu}}$, respectively, the latter possessing eigenvalues $\sigma=0,1$ and eigenprojectors
$\mathbb{P}_{b_{\mu}}^{\sigma}$.

Eq. (\ref{<Q1Q2> A 2}) for the pointer position correlation can be applied to the present case, since the spectra of $\hat{A}$ and $\hat{B}$ used there are allowed to be degenerate. We find
\begin{subequations}
\begin{eqnarray}
\frac{
\langle \hat{Q}_1 \hat{Q}_2 \rangle
^{(
\mathbb{P}_{b_{\mu}} \leftarrow \mathbb{P}_{a_{\nu}}
)}}
{\epsilon_1 \epsilon_2}
&=&
\Re \sum_{\pi, \sigma = 0}^1 \pi \sigma \;
W^{( \mathbb{P}_{b_{\mu}} \leftarrow \mathbb{P}_{a_{\nu}})}
_{\sigma \pi} (\epsilon_1)
\label{<Q1Q2> P 2 a}
\\
&=& \Re W^{( \mathbb{P}_{b_{\mu}} \leftarrow \mathbb{P}_{a_{\nu}})}
_{1 1} (\epsilon_1).
\label{<Q1Q2> P 2 b}
\end{eqnarray}
The quasi-probability (\ref{W(b,a) B A}) reduces in this case to
\begin{eqnarray}
W^{( \mathbb{P}_{b_{\mu}} \leftarrow \mathbb{P}_{a_{\nu}})}
_{\sigma \pi} (\epsilon_1)
= \sum_{\pi'=0}^1
g_{\pi \pi'}^{(\mathbb{P}_{a_{\nu}})}(\epsilon_1)
Tr_s (\rho_{s}
\mathbb{P}^{\pi'}_{a_{\nu}}\mathbb{P}^{\sigma}_{b_{\mu}} \mathbb{P}^{\pi}_{a_{\nu}}
)\; .
\label{<Q1Q2> P 2 c}
\end{eqnarray}
\label{<Q1Q2> P T(b,a)}
\end{subequations}
The decoherence factor is now [see Eq. (\ref{g_nn'_Gauss})]
\begin{equation}
g(\epsilon_1(\pi - \pi')) \equiv g_{\pi \pi'}^{(\mathbb{P}_{a_{\nu}})}(\epsilon_1)
= e^{-\frac{\epsilon_1^2}{8\sigma_{Q_1}^2}(\pi - \pi')^2}.
\label{g_nn' pi}
\end{equation}
Here it is of importance to note that the correlation
$\langle Q_1Q_2 \rangle
^{(
\mathbb{P}_{b_{\mu}} \leftarrow \mathbb{P}_{a_{\nu}}
)}$, Eqs. (\ref{<Q1Q2> P 2 a}), (\ref{<Q1Q2> P 2 b}), is directly proportional to one of the elements of the quasi-probability
(\ref{<Q1Q2> P 2 c}).
Notice also that, for $\epsilon_1 \to \infty$, the quasi-probability
(\ref{<Q1Q2> P 2 c}) reduces to the Wigner rule.

Eqs. (\ref{<Q1Q2> P T(b,a)}) generalize for the successive measurement of two projectors the result (\ref{<Q>proj 1})
for a single measurement.
In contrast to the single-measurement case,
the quasi-probability appearing in
Eq. (\ref{<Q1Q2> P 2 c})
does not coincide, in general, with
the quasi-probability of  $a_{\nu}$ followed by $b_{\mu}$, i.e.,
$W^{(\hat{B} \leftarrow \hat{A})}_{b_{\mu}a_{\nu}}(\epsilon_1)$
of Eq. (\ref{W(b,a) B A}),
since the decoherence factors
$g_{\nu \nu'}^{(\hat{A})}(\epsilon_1)$ and
$g_{1 \pi'}^{(\mathbb{P}_{a_{\nu})}}(\epsilon_1)$
are, in general, different.
This result is a non-classical feature of the successive-measurements problem.

Similarly, when the observables $\mathbb{P}_{a_{\nu}}$ and $\mathbb{P}_{b_{\mu}}$ have been measured in succession, we find the correlation
\begin{equation}
\frac{\langle \hat{P}_1 \hat{Q}_2 \rangle
^{(
\mathbb{P}_{b_{\mu}} \leftarrow \mathbb{P}_{a_{\nu}}
)}}{\epsilon_1 \epsilon_2}
=\frac{1}{2\sigma_{Q_1}^2}
\Im \;
W^{( \mathbb{P}_{b_{\mu}} \leftarrow \mathbb{P}_{a_{\nu}})}_{1 1} (\epsilon_1),
\label{<P1Q2> P e}
\end{equation}
in terms of the same quasi-probability
$W^{( \mathbb{P}_{b_{\mu}} \leftarrow \mathbb{P}_{a_{\nu}})}
_{1 1} (\epsilon_1)$
appearing in Eq. (\ref{<Q1Q2> P 2 b}).

We concentrate on the quasi-probability distribution
$W^{( \mathbb{P}_{b_{\mu}} \leftarrow \mathbb{P}_{a_{\nu}})}_{1 1} (\epsilon_1)$, because it is more directly measurable than that of Eq. (\ref{W(b,a) B A}), in the sense that it can be retrieved directly from the two-pointer correlations when the measurement is performed through the Hamiltonian (\ref{V 2meas}) with
$\hat{A}$ and $\hat{B}$ replaced by projectors.

In Ref. \cite{Johansen-Quantheosuccproj:07} it was shown that from
Kirkwood's quasi-probability distribution one can recover the density matrix $\rho_{s}$ for the system prior to the measurement, provided that the observables are nondegenerate and complementary, i.e., that they have no common eigenprojectors
\cite{Beltrametti+Cassinelli-LogiQuanMech:81}.
We now show that under the same conditions the set
$W^{( \mathbb{P}_{b_{\mu}} \leftarrow \mathbb{P}_{a_{\nu}})}_{1 1} (\epsilon_1)$ {$\forall \mu,\nu$}
also conveys full information about $\rho_{s}$, so that from
successive measurements performed with an arbitrary interaction
strength one could retrieve the full QM state.
From Eq. (\ref{<Q1Q2> P 2 c}) we obtain, writing the projectors in terms of kets and bras,
\begin{eqnarray}
\langle a_{n}|  \rho_{s}   | a_{n'} \rangle
= \frac{1}{G_{n'n}(\epsilon_1)}
\sum_{\mu}
W^{( \mathbb{P}_{b_{\mu}} \leftarrow \mathbb{P}_{a_{n}})}
_{1 1} (\epsilon_1)
\frac{\langle b_{\mu}|a_{n'}\rangle}
{\langle b_{\mu}|a_{n}\rangle}  ,
\label{rho from W(b,a)}
\end{eqnarray}
for $\langle b_{\mu}|a_{n}\rangle \neq 0$.
We have defined [see Eq. (\ref{g_nn' pi})]
\begin{equation}
G_{n n'}(\epsilon_1)
= \delta_{n n'} g^{(P)}_{11}(\epsilon_1)
+(1-\delta_{n n'})g^{(P)}_{10}(\epsilon_1)
=\delta_{n n'} +(1-\delta_{n n'})\exp(-\epsilon_1^2 /8\sigma_{Q_1^2}).
\end{equation}
The result (\ref{rho from W(b,a)}) expresses the matrix elements of the density operator in terms of the quasi-probability distribution
$W^{( \mathbb{P}_{b_{\mu}} \leftarrow \mathbb{P}_{a_{\nu}})}
_{1 1} (\epsilon_1)$.
In the weak-coupling limit,
$\epsilon_1 \to 0$,
$G_{n'n}(\epsilon_1) \to 1$ and
\begin{equation}
\langle a_{n}|  \rho_{s}   | a_{n'} \rangle
\to
 \sum_{\mu} K_{b_{\mu} a_{n}}
  \frac{\langle b_{\mu}|a_{n'}\rangle}
  {\langle b_{\mu}|a_{n}\rangle},
\label{wc rho}
\end{equation}
which coincides with the result found in Eq. (4) of Ref.
\cite{Johansen-Quantheosuccproj:07} in terms of Kirkwood's joint quasi-probability.
In contrast, in the limit $\epsilon \to \infty$
[$G_{n'n}(\epsilon_1) \to \delta_{n' n}$],
\begin{equation}
\langle a_{n}|  \rho_{s}   | a_{n} \rangle
\to
{\sum}_{\mu} W^W_{b_{\mu} a_{n}}
=Tr_s(\rho_{s}\mathbb{P}_{a_n}),
\end{equation}
in terms of Wigner's joint probability.
In this last limit, Eq. (\ref{rho from W(b,a)}) is defined only for $n=n'$: thus only the diagonal elements of the density matrix can be retrieved. Surprisingly, this is precisely the limit in which Wigner's formula is obtained.

We can now express the quantum-mechanical expectation value of an observable in terms of the above quasi-probability distribution.
The expectation value $Tr_s(\rho^{(s)}\hat{O})$ of a Hermitean operator $\hat{O}$ for the system can be written, using Eq. (\ref{rho from W(b,a)}) that relates the $\rho^{(s)}$ matrix elements with the quasi-probability distribution, as
\begin{subequations}
\begin{eqnarray}
Tr_s(\rho^{(s)}\hat{O})
&=& \sum_{n \mu}
W^{( \mathbb{P}_{b_{\mu}} \leftarrow \mathbb{P}_{a_{n}})}
_{1 1} (\epsilon_1) \;
\hat{O}(b_{\mu}, a_n) ;
\hspace{5mm}
\label{<O> 2}
\end{eqnarray}
we have defined the ``transform" of the operator $\hat{O}$ as
\begin{eqnarray}
O(b_{\mu}, a_n)&=&\sum_{n'}
\frac{\langle b_{\mu}|a_{n'}\rangle}
  {\langle b_{\mu}|a_{n}\rangle}
\frac{\langle a_{n'} |  \hat{O} | a_{n} \rangle}
{G_{n' n}(\epsilon_1)} .
\label{O transf}
 \end{eqnarray}
\label{<O> transf}
\end{subequations}
Relations (\ref{<O> transf}) have the structure of a number of transforms found in the literature,
that express the QM expectation value of an observable in terms of its transform and a quasi-probability distribution (in phase space, in the case of Wigner's transform).

\section{Conclusions}

We have demonstrated that the two-pointer correlation in successive measurements may be expressed in terms of a generalized quasi-probability. We found that this quasi-probability reduces to the Wigner formula and to the Kirkwood-Dirac distribution in the strong and weak coupling limit, respectively. We also derived the reduced state after the first measurement, which was found to reduce to L\"uders rule in the strong coupling limit. Furthermore, we found that full information about a quantum system may be obtained from successive measurements of two complementary
observables, regardless of the interaction strength.
This finding, although in resemblance to classical physics, requires the separate measurement of every pair of projectors belonging to the spectral resolution of the two observables. What's more, we even need to ``measure'' not only the position, but also the momentum of the last meter. Thus, reconstructing the state of a system in successive measurements is considerably more complicated in quantum mechanics than in classical physics, where a successive measurement of position and momentum
for every degree of freedom will do the trick.
On the other hand, the quantum scheme is considerably more general
than the classical scheme, since \emph{any} pair of observables may be used to reconstruct a quantum state, provided only that they are nondegenerate and have no common eigenvectors.
The latter condition is equivalent to requiring that the two observables should never possess simultaneous definite values.

We finish with a remark on our particular choice of the pointer states before the measurement as pure Gaussians.
We have shown that when the von Neumann interaction is given by
(\ref{V 2meas}), $\hat{A}$ and $\hat{B}$ being projectors, measurement of the correlation of the
position/momentum of the first pointer with the position of the second, Eqs. (\ref{<Q1Q2> P 2 b}) and (\ref{<P1Q2> P e}),
allows to extract the
real/imaginary part of the quasi-probability
$W^{( \mathbb{P}_{b_{\mu}} \leftarrow \mathbb{P}_{a_{\nu}})}
_{1 1} (\epsilon_1)$.
If the pointers are not described by pure Gaussian states,
those two correlations allow, in general,
to extract the real and imaginary part of two different quasi-probability distributions, which have the same structure, but with different decoherence factors, and merge into a single one for pure Gaussian states.
Either quasi-probability distribution would suffice to retrieve the density matrix for the system.
However, we have not succeeded in devising the appropriate measurable correlations that would allow extracting both the real and imaginary part of either one of the two quasi-probability distributions.

{\ack
PAM is grateful to the Buskerud University College, Norway, LMJ to the C\'atedra Elena Aizen de Moshinsky, Mexico,
and both authors to Conacyt, Mexico (under Contract No. 42655), for
financial support.
PAM wishes to thank M. Bauer for important discussions on the present paper.
}


\begin{thebibliography}{10}

\bibitem{Neumann-MathFounQuanMech:55}
J.~{von Neumann}, Mathematical Foundations of Quantum Mechanics, Princeton
  University Press, Princeton, 1955.

\bibitem{Wigner-ProbMeas:63}
E.~P. Wigner, Am. J. Phys. 31 (1963) 6--15.

\bibitem{Lueders-UEbeZust:51}
G.~L\"{u}ders, Ann. Physik 8~(6) (1951) 322--328.

\bibitem{Johansen-Quantheosuccproj:07}
L.~M. Johansen, Phys. Rev. A 76 (2007) 012119.

\bibitem{Bell+Nauenberg-MoraAspeQuanMech:66}
J.~S. Bell, M.~Nauenberg, in: A.~de~Shalit, H.~Feshbach, L.~van~der Hove (Eds.), Preludes in Theoretical
  Physics, North-Holland Publishing Company, Amsterdam, 1966, pp. 279--286.
  Reprinted in J.~S. Bell, Speakable and unspeakable in quantum mechanics, Cambridge
  University Press, Cambridge, 1987, Ch.~3, pp. 22--28.

\bibitem{Arthurs+Kelly-SimuMeasPairConj:65}
E.~Arthurs, J.~L. {Kelly, Jr.}, Bell Syst. Tech. J. 44 (1965) 725--729.

\bibitem{Aharonov+AlbertETAL-ResuMeasCompSpin:88}
Y.~Aharonov, D.~Z. Albert, L.~Vaidman, Phys. Rev. Lett. 60~(14) (1988) 1351--1354.

\bibitem{Peres-QuanLimiDeteWeak:89}
A.~Peres, Phys. Rev. D 39~(10) (1989) 2943--2950.

\bibitem{Lamb-Sequmeasquanmech:87}
W.~E. Lamb, in: E.~R. Pike,  S.~Sarkar (Eds.), Quantum Measurement and Chaos, Plenum Press, 1987, pp.
  183--193.

\bibitem{Hay+Peres-QuanClasDescMeas:98}
O.~Hay, A.~Peres, Phys. Rev. A 58~(1) (1998) 116--122.

\bibitem{Kraus-StatEffeOper:83}
K.~Kraus, States, Effects and Operations, no. 190 in Lecture Notes in Physics, Springer Verlag, Berlin, 1983.

\bibitem{konrad}
T. Konrad,
Less is More-On the Theory and Application of Weak and Unsharp Measurements in Quantum Mechanics,
Ph. D. Dissertation, University of Konstanz, 2003.

\bibitem{Kirkwood-QuanStatAlmoClas:33}
J.~G. Kirkwood, Phys. Rev. 44 (1933) 31--37.

\bibitem{Dirac-AnalBetwClasQuan:45}
P.~A.~M. Dirac, Rev. Mod. Phys. 17~(2/3) (1945) 195--199.

\bibitem{Ludwig-Mess:53}
G.~Ludwig, Z. Phys. 135 (1953) 483--511.

\bibitem{Goldberger+Watson-MeasTimeCorrQuan:64}
M.~L. Goldberger, K.~M. Watson, Phys. Rev. 134~(4B) (1964) B919--B928.

\bibitem{Steinberg-Condprobquantheo:95}
A.~M. Steinberg, Phys. Rev. A 52~(1) (1995) 32--42.

\bibitem{Johansen+Luis-NoncWeakMeas:04}
L.~M. Johansen, A.~Luis, Phys. Rev. A 70 (2004) 052115.

\bibitem{Margenau+Hill-CorrbetwMeasQuan:61}
H.~Margenau, R.~N. Hill, Prog. Theor. Phys. 26~(5) (1961) 722--738.

\bibitem{Beltrametti+Cassinelli-LogiQuanMech:81}
E.~G. Beltrametti, G.~Cassinelli, The Logic of Quantum Mechanics,
  Addison-Wesley Publishing Company, Massachusetts, 1981.

\end{thebibliography}
\end{document}